# Quasi-Ballistic Thermal Conduction in 6H-SiC

Zhe Cheng[1,a)], Weifang Lu[2,a),*], Jingjing Shi[1,a)], Daiki Tanaka[2], Nakib H. Protik[3], Shangkun Wang[1], Motoaki Iwaya[2], Tetsuya Takeuchi[2], Satoshi Kamiyama[2], Isamu Akasaki[2,5], Hiroshi Amano[4], Samuel Graham[1,*]

[1] George W. Woodruff School of Mechanical Engineering, Georgia Institute of Technology, Atlanta, Georgia 30332, USA

[2] Department of Materials Science and Engineering, Meijo University, 1-501 Shiogamaguchi, Tenpaku-ku, Nagoya 468-8502, Japan

[3] John A. Paulson School of Engineering and Applied Sciences, Harvard University, Cambridge, Massachusetts 02138, USA

[4] Institute of Materials and Systems for Sustainability, Nagoya University, Furo-cho, Chikusa-ku, Nagoya 464-8601, Japan

[5] Akasaki Research Center, Nagoya University, Furo-cho, Chikusaku-ku, Nagoya 460-8601, Japan

[a)]These authors contributed equally

[*]Corresponding authors: weif@meijo-u.ac.jp; sgraham@gatech.edu

# ABSTRACT

The minimization of electronics makes heat dissipation of related devices an increasing challenge. When the size of materials is smaller than the phonon mean free paths, phonons transport without internal scatterings and laws of diffusive thermal conduction fail, resulting in significant reduction in the effective thermal conductivity. This work reports, for the first time, the temperature dependent thermal conductivity of doped epitaxial 6H-SiC and monocrystalline porous 6H-SiC below room temperature probed by time-domain thermoreflectance. Strong quasi-ballistic thermal transport was observed in these samples, especially at low temperatures. Doping and structural boundaries were applied to tune the quasi-ballistic thermal transport since dopants selectively scatter high-frequency phonons while boundaries scatter phonons with long mean free paths. Exceptionally strong phonon scattering by boron dopants are observed, compared to nitrogen dopants. Furthermore, orders of magnitude reduction in the measured thermal conductivity was observed at low temperatures for the porous 6H-SiC compared to the epitaxial 6H-SiC. Finally, first principles calculations and a simple Callaway model are built to understand the measured thermal conductivities. Our work sheds light on the fundamental understanding of thermal conduction in technologically-important wide bandgap semiconductors such as 6H-SiC and will impact applications such as thermal management of 6H-SiC-related electronics and devices.

# INTRODUCTION

Silicon Carbide (SiC) is a wide bandgap semiconductor which has been widely used in power electronics and optoelectronics.[1] Due to its high thermal conductivity and relatively small lattice mismatch with Gallium Nitride (GaN), SiC is also used as substrates for applications such as GaN-on-SiC power devices.[2] The Joule-heating in these devices results in localized hotspots which degrades device performance and reliability, while the understanding of heat conduction mechanisms in this technologically-important material is still limited.[3-6] When the size of the material is smaller than the phonon mean free paths, the phonons that carry the thermal energy travel in the material without internal scattering events; this transport regime is called ballistic thermal transport.[7-9] Thermal physics laws which are based on the diffusion assumption such as Fourier's law breaks down under such transport conditions since the thermal conductivity becomes a material-size dependent property and the definition of local temperature becomes difficult due to thermal non-equilibrium.[7-9] The ballistic thermal transport phenomena are increasingly related to real-world electronics applications with the minimization of electronics. The shrinking sizes of material dimensions and Joule-heating sources reduce the thermal conductivity of device components, which makes heat dissipation in these devices an increasing challenge.

Most of the previous thermal studies on SiC focused on the measurements and predictions of the thermal conductivity of bulk materials, especially for 6H-SiC.[3-6] The experimental measurements of thermal conductivity of 6H-SiC only have data at high temperatures while low temperature data of thermal conductivity is not available despite the effect of ballistic thermal transport increases with decreasing temperature. At low temperatures, the number of excited phonons decreases, leading to lower phonon scattering rates. As a result, the phonon mean free path increases, leading to increasingly strong ballistic thermal transport. Additionally, recent calculations showed that the boron (B) dopants in 3C-SiC induce exceptionally strong phonon scattering.[6] Unlike structural boundaries which scatters phonons with long mean free paths, substitutions of doping atoms are more likely to scatter high-frequency phonons. How these doping effects tune the ballistic thermal transport is still an open question, especially at low temperatures when doping concentrations affect thermal conductivity significantly.

The development of time-domain thermoreflectance (TDTR) provides a solution to probe ballistic thermal transport by controlling the laser spot size.[10] A pump beam heats up the sample surface periodically while a probe beam detects the temperature of the sample surface via thermoreflectance. Phonons with mean free paths longer than the beam size do not contribute to the measured effective thermal conductivity. Only the thermal conductivity contributed by the phonons with mean free paths shorter than the beam size is measured.[10] By measuring the effective thermal conductivity, the effect of ballistic thermal transport can be probed under variable conditions, for instance, temperature, doping types, doping concentrations, and structural boundaries.

In this work, we measure the temperature-dependent effective thermal conductivity of epitaxial monocrystalline 6H-SiC doped by B and nitrogen (N) and porous monocrystalline 6H-SiC by TDTR. Dopants selectively scatter high-frequency phonons while structural boundaries scatter phonons with long mean free paths. Density functional theory (DFT) based first principles calculations are performed to calculate thermal conductivity and phonon mean free paths in the perfect single crystal of 6H-SiC. A simple Callaway mode is also built to understand the measured thermal conductivity.

**RESULTS AND DISCUSSIONS**

In this work, four 6H-SiC samples were fabricated and measured. Three homo-epitaxial 6H-SiC films (Sample A, Sample B, and Sample C) with different doping types and concentrations were grown on commercially-available 6H-SiC substrates (TankeBlue, Beijing). All three as-grown films are thicker than 100 µm. More details about the sample growth process can be found in the Experimental Section. The fourth sample is a porous 6H-SiC fabricated by electrochemical etching which will be discussed in detail later. Figure 1 shows the structures of the three epitaxial 6H-SiC samples. As shown in Figure 1(a), under the illumination of a 325-nm laser, the epitaxial SiC layer of Sample B emits yellowish fluorescence due to the donor-acceptor-pair recombination.[11] The epi-layer thickness can be determined according to these cross-sectional microscopy images. The epi-layer thicknesses are 113 µm for Sample A, 128 µm for Sample B, and 125 µm for Sample C. Figure 1(b) shows the top surface of Sample B before and after molten KOH etching at 540 °C for 10 min. KOH etching is applied to extend the dislocations, making them visible under optical microscopy and enabling the quantification of dislocation density. The estimated threading edge dislocation densities are $6.67 \times 10^3$ cm$^{-2}$ for Sample A, $4.58 \times 10^4$ cm$^{-2}$ for Sample B, and $4.27 \times 10^4$ cm$^{-2}$ for Sample C. These low dislocation densities are not expected to affect thermal conductivity.[12] Figures 1(c-d) show the boron and nitrogen doping concentration distributions along the thickness direction measured by secondary ion mass spectrometry (SIMS). About 50% variations of the doping concentrations near the sample surface and uniform doping concentrations at areas 1-2 µm under the sample surface are observed. These thin layers with doping concentration variations near the sample surface are due to unintentional epi-growth. When the scheduled growth process was completed, the supply of Ar and $N_2$ gas was turned off. Meanwhile, the precursors remained in the chamber and the sublimation continued until the temperature completed its ramp down.

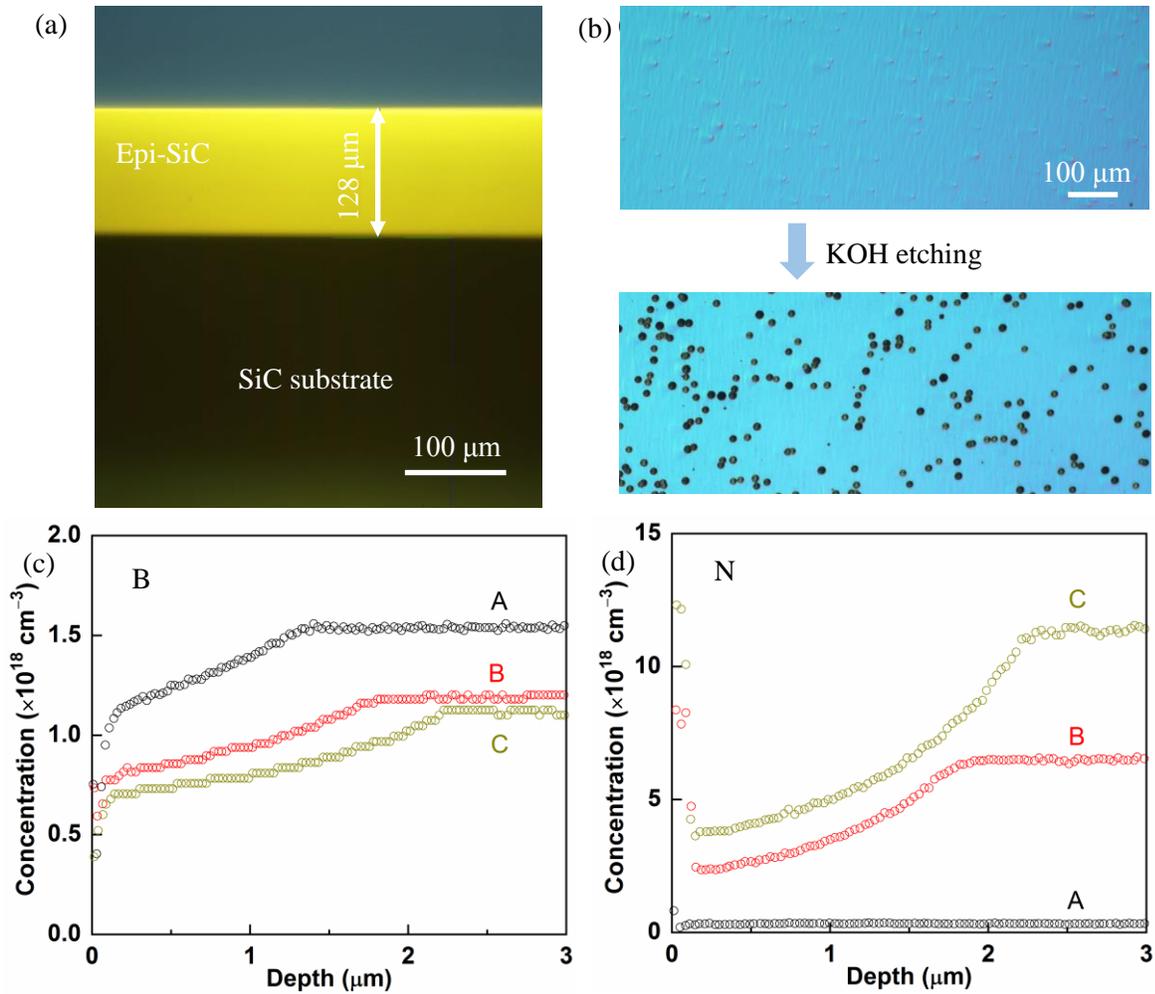

Figure 1. The structures of the three epitaxial 6H-SiC samples. (a) A typical cross-sectional microscope image of Sample B under the illumination of 325 nm laser, which shows the fluorescent SiC epi-layer. (b) The top surface of Sample B before and after KOH etching. (c) B

doping concentration distributions along the thickness direction measured by SIMS. (d) N doping concentration distribution along the thickness direction measured by SIMS.

The temperature dependent thermal conductivity of Sample A, Sample B, and Sample C were measured by TDTR with a 10× objective and a modulation frequency of 3.6 MHz. A layer of Al is deposited on all the sample surface as transducer. The Al thickness was determined picosecond acoustic as 160 nm in this work. Figure 2(a) shows the temperature dependent thermal conductivity of Sample A, Sample B, and Sample C. The cross-plane direction of these 6H-SiC samples are parallel to the c-axis so the measured thermal conductivity is the cross-plane thermal conductivity documented in the literature. It is notable that the cross-plane thermal conductivity of 6H-SiC at low temperatures has not been reported before. The thermal conductivities obtained from a full solution of the phonon BTE without considering finite size effects are shown in solid lines. At room temperature and above, the measured thermal conductivities are close to other measured thermal conductivities in the literature and agree well with the first principles calculated values.[3-5,13] However, at low temperatures, the measured thermal conductivities of Sample A, Sample B, and Sample C are significantly reduced compared with calculated thermal conductivity. The low temperature discrepancy can be understood by considering the effects of the impurity scattering and the cross-over to the quasi-ballistic transport regime.

The phonon mean free path accumulated thermal conductivity obtained from the BTE at different temperatures are shown in Figure 2(b). At 300 K, 80% of thermal conductivity is contributed by phonons with mean free path shorter than 4 μm. When the temperature decreases, the phonon mean free path increases due to the less extensive phonon-phonon scattering. In TDTR measurements, phonons with mean free path longer than the laser beam size transport directly out of the heating region and do not contribute to the measured thermal conductivity.[10] TDTR does not detect long wavelength phonons whose mean free paths are longer than the spot size. This quasi-ballistic thermal transport enables the probing of the contribution of phonons with mean free paths smaller than a certain spot size to thermal conductivity.[10,14,15] The pump and probe diameters are 21.3 μm and 11.6 μm, respectively. The root-mean-square spot size (radius) is 8.6 μm, which is larger than 4 μm. As a result, phonons which contribute to the majority of the thermal conductivity of 6H-SiC at room temperature transport diffusively. This explains why the TDTR-measured thermal conductivity matches well with the first principles calculated values for the perfect single crystalline SiC.

At low temperatures, the mean free paths of phonons which contribute to the majority of the thermal conductivity increases and become larger than the spot size. The effect of quasi-ballistic thermal conduction becomes significant and leads to the reduction of measured thermal conductivity. As shown in Figure 2(c), the lines are calculated thermal conductivity contributed by phonons with mean free paths shorter than 20 μm and 7.9 μm. The measured thermal conductivities of Sample A, Sample B, and Sample C at low temperatures are located between these two calculated lines. By considering the spot size of the TDTR laser, the measured thermal conductivity can be well-explained by the quasi-ballistic thermal conduction, which confirms the ability of TDTR to probe non-diffusive thermal transport.

By tuning the doping elements and doping concentrations, the measured thermal conductivity is tuned at low temperatures, as shown in Figures 2(c-d). At low temperatures, phonon modes with high scattering rates are not thermally activated, and the phonon-defect scattering dominates over the phonon-phonon scattering, making the former scattering channel limit the mean free paths. By measuring thermal conductivity at low temperatures, we are able to probe the effects of different doping types on quasi-ballistic thermal conduction and study how they selectively scatter phonons. As shown in Figure 2(d), we see obvious differences in the measured thermal conductivity at low temperatures for samples with different dopants. The dash line shows the calculated thermal conductivity which only considers the quasi-ballistic thermal conduction limited by a spot-size of about 10 μm. The deviations in thermal conductivity below 130 K is attributed to the effect of dopants. Point defects prefer to scatter high-frequency phonons.[6] Quasi-ballistic thermal transport makes TDTR measurements unable to detect the contribution of phonons with long mean free paths. Phonons with long mean free paths are less likely scattered by dopants. In this scenario, by removing the contributions of phonons with long mean free paths, the measured thermal conductivity is affected more significantly by the dopants, compared to the condition without removing the contributions of phonons with long mean free paths. Therefore, quasi-ballistic thermal conduction in the TDTR measurements can be tuned relatively more significantly by tuning the doping types and concentrations after removing the contributions of phonons with long mean free path.

The samples are co-doped by boron and nitrogen. The mass differences are the same for carbon substituted by boron and nitrogen, but the measured thermal conductivity is significantly affected by the boron doping concentration for the three samples in this work. Sample A has the smallest total doping concentration but has the highest boron doping. The measured thermal conductivity of Sample A is the lowest at low temperatures. For Sample B and Sample C, they have a similar concentration of boron doping. Sample C has a higher concentration of nitrogen doping, resulting in a lower thermal conductivity compared with Sample B at low temperatures. This is consistent with the calculations for 3C-SiC in literature that boron doping has a much stronger effect than any other defect types.[6] It is worth noting that according to the classical mass-difference model of impurity scattering, the effect of boron and nitrogen should be the same. This highlights the limitations of the simple mass-difference models in capturing the real effect of the various substitution defects. The exceptionally strong phonon scattering by boron due to resonant phonon scattering should apply to 6H-SiC as well.

The solid lines in Figure 2(d) show the calculated thermal conductivity which consider both quasi-ballistic thermal conduction and dopant scatterings. The dopant scattering is included according to a simple Callaway model. More details can be found in the Methods section. We use 10 μm here in the calculation for the spot size. Phonons with mean free paths longer than 10 μm do not contribute to the calculated thermal conductivity. The solid lines agree with the measured thermal conductivity better than the dashed line, which confirms that defect scattering reduces part of the measured thermal conductivity. The difference among these solid lines are relatively small according to the used formula of defect scattering rate while the difference in the measured thermal conductivity are obvious. These deviations are possibly due to unusual phonon scattering by boron in 6H-SiC and possible defect complexes.[6,16-19]

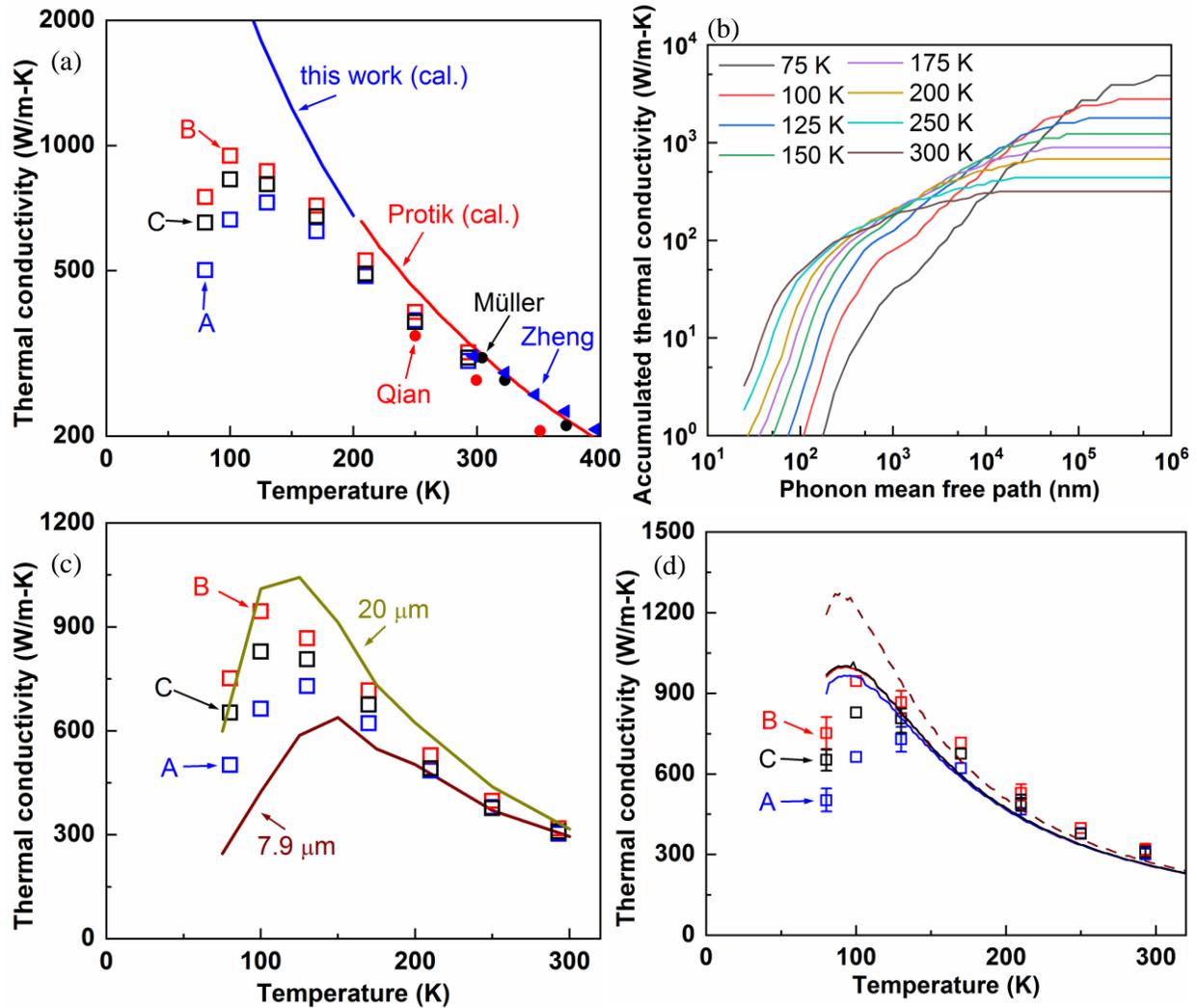

Figure 2. Thermal conductivity of 6H-SiC. (a) TDTR-measured thermal conductivity of Sample A, Sample B, and Sample C compared with experimentally measured values and calculated values in the literature.[3-5,13] (b) First principles calculated phonon mean free path accumulated thermal conductivity of 6H SiC at different temperatures. (c) Temperature dependence of measured thermal conductivity compared with first principles calculated thermal conductivity contributed by phonons with mean free path shorter than 20 μm and 7.9 μm without considering doping effects. (d) Temperature dependence of measured thermal conductivity compared with calculated thermal conductivity by a simple Callaway model which considers only quasi-ballistic thermal transport

(dash line) and both quasi-ballistic thermal transport and phonon scatterings by dopants (solid lines). The spot size used in the calculation is 10 μm.

Dopants prefer to scatter high-frequency phonons which usually have relatively short mean free paths, while boundaries scatter phonons with long mean free paths. To fabricate a monocrystalline sample with boundaries, we etched a n-type 6H-SiC sample to obtain nanoscale porous structures to scatter phonons with long mean free path. Figure 3(a) shows the SEM images of Sample D. The left one is the surface of the etched sample while the right one is the cross-section of the etched porous structure. We apply a double-frequency TDTR measurement to obtain the thermal conductivity and density simultaneously. Figures 3(b-c) show the sensitivity of the porous SiC thermal conductivity, porous SiC density, and Al-SiC thermal boundary conductance (TBC) with modulation frequency of 3.6 MHz and 1.2 MHz, respectively. The sensitivity of measuring a certain unknown parameter by TDTR is defined as

$$S_i = \frac{\partial \ln(-V_{in}/V_{out})}{\partial \ln(p_i)}, \quad (1)$$

where $S_i$ is the sensitivity to parameter $i$, $-V_{in}/V_{out}$ is the TDTR ratio, $p_i$ is the value of parameter $i$. For the same spot on a sample, different modulation frequencies result in different thermal penetration depths and different sensitivities for a certain unknown parameter. As shown in Figures 3(b-c), the sensitivity of SiC thermal conductivity with modulation frequency of 1.2 MHz is much larger than the SiC density while it is comparable with the SiC density with modulation frequency of 3.6 MHz. The difference in sensitivities helps separate the two unknown parameters. Figure 3(d) shows the excellent agreement between experimental data and theoretical fittings. The data with different modulation frequencies were fitted simultaneously to obtain the SiC thermal conductivity and the SiC density.

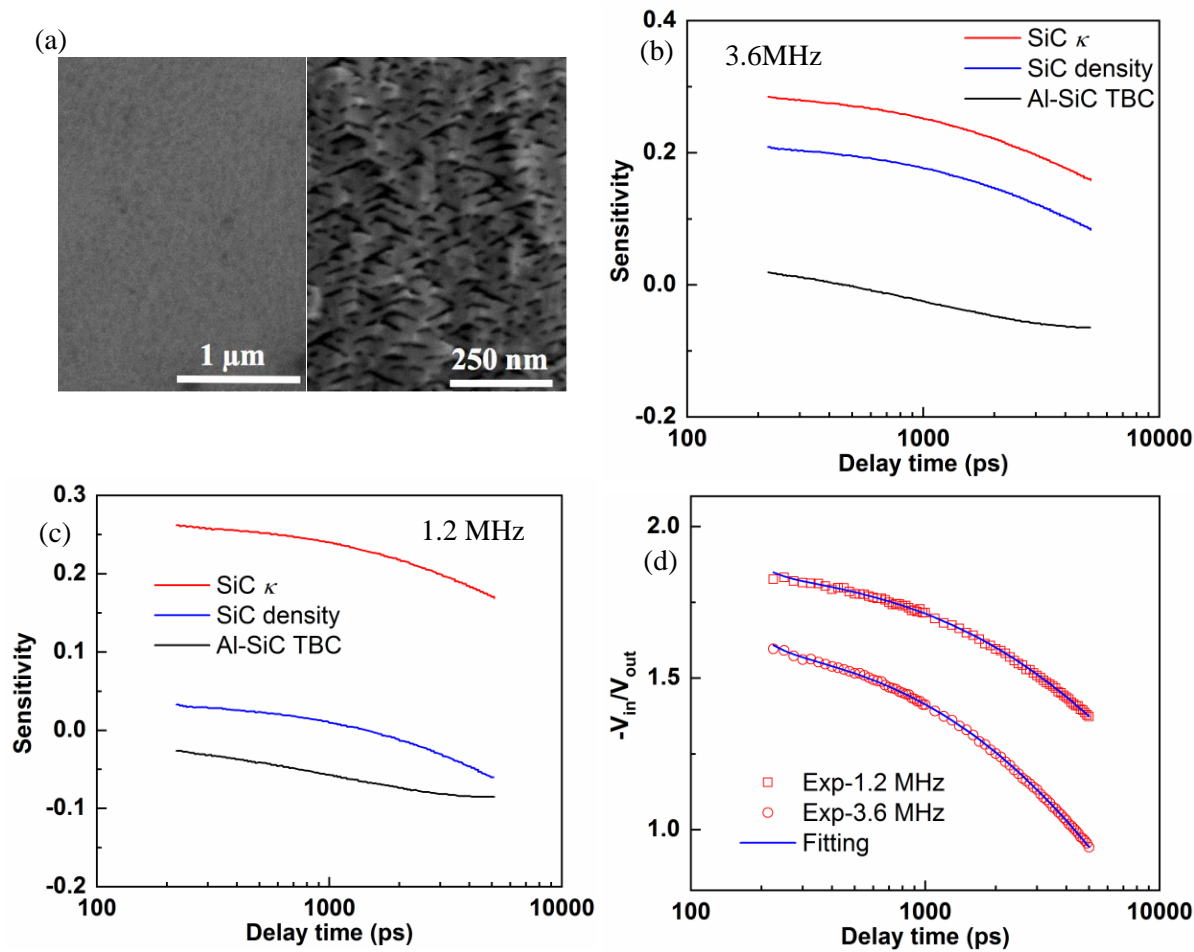

Figure 3. Thermal characterizations of Sample D. (a) surface (left) and cross-section (right) SEM images of Sample D. (b) TDTR sensitivity of unknown parameters with modulation frequency of 3.6 MHz. (c) TDTR sensitivity of unknown parameters with modulation frequency of 1.2 MHz. (d) Excellent fitting of experimental and theoretical data.

The measured thermal conductivity of Sample D is shown in Figure 4. The solid volume fraction (*P*) is about 30% (porosity is 70%). The temperature dependent thermal conductivity of Sample B is also included as comparison since they have similar dopant concentrations. The thermal conductivity of solid SiC estimated by the solid volume fraction and the effective medium theory expression are also included in Figure 4. The formula for solid volume fraction is:[9]

$$k_{\text{solid,p}} = k_{\text{eff}}/P, \qquad (2)$$

The Maxwell effective medium theory applies for conduction in a solid matrix that contains a uniform but dilute arrangement of spherical pores:[20,21]

$$k_{\text{solid,M}} = k_{\text{eff}}/\left(\frac{2P}{3-P}\right), \qquad (3)$$

Both the solid volume fraction and Maxwell effective medium theory formula are based on diffusive heat conduction assumption and do not consider detailed phonon scattering mechanisms. Here, we use it to estimate and remove the effect of pore volume on thermal conductivity.

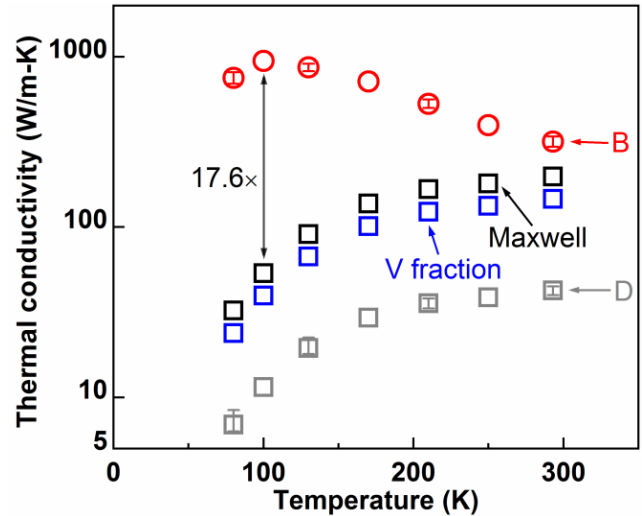

Figure 4. Temperature dependence of the measured thermal conductivity and the calculated solid thermal conductivity of Sample D.

Compared with the measured thermal conductivity of Sample B, the solid thermal conductivity of Sample D is 17.6 times smaller at 100 K. Even at room temperature, the solid thermal conductivity of Sample D is much lower than that of Sample B. This is attributed to the strong ballistic thermal transport in porous SiC, even at room temperature. The pores create the additional boundaries. When phonons with mean free paths longer than the structural length limited by these boundaries, phonons scatter with the boundaries without any internal scattering with other phonons. As temperature decreases, the phonon mean free paths increases. The structural length does not change with temperature, leading to an increased number of phonons which transport without internal phonon-phonon scattering. As a result, we see an increased difference between the thermal conductivity of Sample B and Sample D. Since the pore size is much smaller than the spot size of TDTR laser, we do not expect strong quasi-ballistic thermal conduction from TDTR measurements as the epitaxial samples. Compared to the effects of spot size and doping, the reduction in thermal conductivity caused by boundaries-induced quasi-ballistic thermal transport are much more significant, especially at low temperatures.

## CONCLUSIONS

This work reported the temperature dependence of thermal conductivity of doped epitaxial 6H-SiC and monocrystalline porous 6H-SiC probed by TDTR. Quasi-ballistic thermal transport in these 6H-SiC samples was observed, especially at low temperatures. Doping was used to tune the quasi-ballistic thermal transport by selectively scattering high-frequency phonons. It is found that B dopants in 6H-SiC scatter phonons much more significantly than N dopants. Additionally, boundaries created by the porous structure scatter phonons with long mean free paths. Orders of magnitude reduction in thermal conductivity was observed for the porous 6H-SiC compared to the epitaxial 6H-SiC at low temperatures, indicating strong quasi-ballistic thermal conductivity. First principles calculations and a simple Callaway mode were built to understand the measured thermal conductivity. Our work sheds light on the heat conduction mechanisms in 6H-SiC and will impact applications of SiC-based electronics since the sizes of these devices become comparable to the phonon mean free paths in SiC. In this scenario, quasi-ballistic thermal transport plays an important role. Furthermore, the thermal conductivity could be orders of magnitude lower than the bulk value, possibly leading to overheating of electronics and degrading the devices.

**EXPERIMENTAL SECTION**

**Epitaxial Growth of Fluorescent 6H-SiC.** The fluorescent SiC samples were grown in a quasi-closed carbon crucible via a closed sublimation technique.[22-24] The crucible reactor consists of two parts for crystal growth and evaporation of the doping source, respectively, wherein the boron (B) and nitrogen (N) doping sources were arranged at the bottom. Commercial n-type 6H-SiC substrates (TankeBlue, Beijing) with a miscut angle of 3.5° toward the <११2̲0> direction were used as the seed substrates. A polycrystalline 3C-SiC plate (Ferrotec Material Technologies Corporation, Tokyo, Japan) was used as the source material. Si-terminated surface of the 6H-SiC substrate was exposed toward the source material for epitaxial growth. Specifically, the distance between the source and the seed substrate was set to 1 mm, where the growth process was driven by the temperature gradient between the source and the seed substrate. To prevent the carbonization and improve the SiC crystalline quality, several tantalum (Ta) components were employed near the source substrates for absorbing carbon. During epitaxial growth, B and N doping was mainly induced using BN powder, while the doping concentration in samples A, B, and C was tuned under ambient Argon, 50-sccm $N_2$, and 100-sccm $N_2$, respectively. Here, we carried out a dummy growth with a constant amount of BN powder prior to the 6H-SiC growth; then the B-N atoms can stick or be absorbed inside the crucible. Using such kind of "memory effect", B and N can be easily incorporated during epi-growth. To establish a fast growth rate (around 80 µm/h), quasi-closed spaces were heated to a stable temperature of 1900 ºC for crystal growth. Sample A is p-type doping while Sample B and Sample C are n-type doping. The growth rates for samples A, B, and C were 79.6, 88.6, and 85.2 µm/h, respectively.

**Porous SiC Fabrication.** The porous SiC sample is fabricated by an anodic oxidation etching technique.[25,26] The doping concentrations of the SiC substrates, the used current density and voltage, and the electrolyte (hydrofluoric acid) concentrations jointly affect the porous morphology during the anodic oxidation etching process.[25,26] In this work, the porous SiC was fabricated on a commercial n-type 6H-SiC substrate with background doing level of B ($5.37 \times 10^{18}$ $cm^{-3}$) and N ($0.95 \times 10^{18}$ $cm^{-3}$). To achieve an ohmic contact, a 100 nm-thick nickel layer followed by a 10 nm-thick titanium layer and a 200 nm-thick gold film was deposited on the back side of the SiC substrate using an electron-beam evaporation system (Ei-5, ULVAC Co., USA). Afterwards, the anodic oxidation process was carried out in a Teflon cell with a platinum cathode, wherein the backside was connected to an anode via conductive tapes. During the anodic oxidation, the SiC substrate was immersed in aqueous 5 wt. % HF solution for 2 hours under a pulsed-constant voltage of 6 V. To improve the uniformity of pores, UV illumination (365 nm) was simultaneously performed to facilitate the penetration of holes into deep layer. A 27 µm-thick porous layer was formed on the SiC substrate, confirmed by a scanning electron microscopy (SEM, SU70, Hitachi High Technologies Co., Japan). Uniform dendritic porous structures were observed through the entire porous layer.

**Thermal Characterization.** TDTR is an optical pump and probe technique to measure thermal properties of both nanostructures and bulk materials.[27,28] A pump beam (400 nm) chopped by an electro-optical modulator heats the sample surface periodically while a delayed probe beam (800 nm) detects the temperature of the sample surface.[29] The signal picked up by a photodetector and a lock-in amplifier is fitted with an analytical heat transfer solution of the sample structure to infer unknown parameters.[29] By changing the modulation frequency, the thermal penetration depth can be tuned to obtain different sensitivity of each unknown parameter. By measuring the same spot with multiple modulation frequencies, multiple unknown thermal properties can be extracted simultaneously.[28] In the data fitting, the Al thermal conductivity is obtained by measuring its electrical conductivity and applying Wiedemann-Franz law. The heat capacity of Al and SiC are from literature.[3] The error bars are calculated based a Monte Carlo method.[30]

**First principles Calculations.** The first principles calculations of bulk 6H-SiC thermal conductivity follow the calculation process described in Ref.[5] Density functional theory (DFT) and density functional perturbation theory (DFPT) results are obtained from Ref.[5] For low temperature thermal conductivity calculation, the ShengBTE software[31] is applied to calculate the thermal conductivity tensor using a locally adaptive Gaussian broadening method to approximate the energy conserving delta functions appearing in the scattering rates expressions, and the detailed q-mesh information can be found in Ref.[5] With the ShengBTE software, phonon mean free path accumulated thermal conductivity tensors from 75 to 200 K are achieved.

**Callaway Model.** With the phonon dispersion relation, the expression of thermal conductivity $k$ in the form of frequency space integration based on the Boltzmann transport equation (BTE) is[32,33]:

$$k = \frac{1}{3}\Sigma_p \int_0^{\omega_{\text{cut-off}}} \hbar\omega D_\lambda \frac{df_{BE}}{dT} v_\lambda^2 \tau_{C,\lambda}, \qquad (4)$$

where $\Sigma_p$ is over all phonon polarizations (36 for 6H-SiC), and $D_\lambda$ is the modal phonon density of states, $f_{BE}$ is the Bose-Einstein distribution function, the 1/3 comes from the isotropic assumption, $\lambda$ is phonon mode, $v_\lambda$ is the modal phonon group velocity, and $\tau_{C,\lambda}$ is the modal combined relaxation time. The combined relaxation time $\tau_{C,\lambda}$ of each phonon mode can be obtained from the Matthiessen's rule as[32,33]:

$$\tau_{C,\lambda} = \left(\frac{1}{\tau_U} + \frac{1}{\tau_M} + \frac{1}{\tau_B}\right)^{-1}, \qquad (5)$$

where $\tau_U$, $\tau_M$, and $\tau_B$ are the relaxation time of Umklapp phonon-phonon scattering, mass-difference phonon-impurity scattering, and phonon-boundary scattering, respectively, and the scattering rate expressions are[34,35]:

$$\frac{1}{\tau_U} = BT\omega^2 e^{-\frac{C}{T}}, \qquad (6)$$

$$\frac{1}{\tau_M} = \frac{V\omega^4}{4\pi v^3}\Sigma_i x_i \left(\frac{\Delta M_i}{M}\right)^2, \qquad (7)$$

$$\frac{1}{\tau_B} = v/d, \qquad (8)$$

where $B$ and $C$ are fitting parameters, $V$ is the volume of the 6H-SiC primitive cell, $x_i$ is the atomic fraction of sites occupied by defect $i$, $\Delta M_i$ is the mass difference between defect and original atom, and d is the thickness of the SiC sample.

When calculating the scattering rates of different mechanisms, the parameters $B$ and $C$ are first optimized to have the best agreement with the SiC thermal conductivity predicted from the first principles calculations. Here the thermal conductivity data agrees well with DFT results from 75 to 400 K to make sure that the thermal conductivity related phonon-phonon scattering is accurately captured in the model. After carefully fitting B and C for the phonon-phonon scattering rates, the impurity scattering rates of boron and nitrogen are calculated from the SIMS results. Nitrogen defects are at the carbon sites while boron atoms can replace either carbon or silicon sites, depending on the growth conditions.[36] In carbon-rich growth conditions, boron replaces silicon while in silicon-rich conditions, boron replaces carbon.[36,37] Here, we assume boron replace carbon in our calculations since our silicon-rich growth condition, similar to the calculations for 3C-SiC in the literature.[6] As mentioned above, even with similar mass difference, the scattering rate of boron and nitrogen differs significantly.[6] Here we assume that the scattering rate ratio of 6H-SiC of boron over isotope with the same concentration is the same as that of 3C-SiC, and we applied the scattering information in Ref.[6] to our calculation.

To consider the quasi-ballistic thermal transport in the Callaway model, for each phonon mode in the calculation, the mean free path is taken as the modal phonon group velocity times the relaxation time. The relaxation time is the multiplicative inverse of the total scattering rate (a combination of phonon-phonon scattering, defect scattering, and boundary scattering using the Matthiessen rule). We only include the thermal conductivity contribution from phonon modes with mean free path smaller than the spot size to consider the effect of quasi-ballistic thermal transport.

**Supporting Information:** The supporting information includes the temperature dependence of the measured Al-SiC TBC for Sample A, Sample B, and Sample C (Figure S1), and the fabrication process of Sample D and schematic diagram of pore structure (Figure S2).

**Competing interests:** The authors claim no competing financial interests.

**ACKNOWLEDGEMENT**: Z.C. and J. S. would like to thank Prof. Tianli Feng for helpful discussions. Z. C., J. S., and S. G. would like to acknowledge the financial support from U.S. Office of Naval Research under a MURI program (Grant No. N00014-18-1-2429). W. L., D. T., M. I., T. T., S. K., and I. A. would like to acknowledge MEXT "Program for research and development of next-generation semiconductor to realize energy-saving society" [No. JPJ005357], MEXT "Private University Research Branding Project", JSPS KAKENHI for Scientific Research A [No.15H02019], JSPS KAKENHI for Scientific Research A [No.17H01055], JSPS KAKENHI for Innovative Areas [No.16H06416], and Japan Science and Technology CREST [No.16815710].